\documentclass[aps,prd,preprint,superscriptaddress,tightenlines,nofootinbib]{revtex4}

\usepackage{graphicx}% Include figure files
\usepackage{dcolumn}% Align table columns on decimal point
\usepackage{bm}% bold math

\newcommand{\ccbar}{c \bar c}
\newcommand{\gev}{\,\mbox{GeV}}
\newcommand{\mev}{\,\mbox{MeV}}
\newcommand{\invpb}{\,\mbox{pb}^{-1}}
\newcommand{\dilep}{\ell^+\ell^-}
\newcommand{\diel}{e^+e^-}
\newcommand{\dimu}{\mu^+\mu^-}
\newcommand{\jpsi}{J/\psi}
\newcommand{\pp}{\psi(2S)}
\newcommand{\chicJ}{\chi_{cJ}}
\newcommand{\egammalow}{E_{\gamma\mbox{-low}}}
\newcommand{\jpsiplusany}{X\jpsi}
\newcommand{\PiPiJ}{$\pi^+\pi^-\jpsi$}
\newcommand{\PizPizJ}{$\pi^0\pi^0\jpsi$}
\newcommand{\PizJ}{$\pi^0\jpsi$}

\newcommand{\EtaJGG}{$\eta(\to \gamma\gamma)\jpsi$}
\newcommand{\EtaJThreePi}{$\eta(\to \pi^+\pi^-\pi^0)\jpsi$}
\newcommand{\XJ}{$X\jpsi$}

\newcommand{\ChicZero}{$\gamma \chi_{c0} \to \gamma\gamma\jpsi$}
\newcommand{\ChicOne}{$\gamma \chi_{c1} \to \gamma\gamma\jpsi$}
\newcommand{\ChicTwo}{$\gamma \chi_{c2} \to \gamma\gamma\jpsi$}

\begin{document}
\preprint{CLNS 05/1909}       % for CLNS notes
\preprint{CLEO 05-4}         % for CLNS notes

\title{Branching Fractions for $\psi(2S)$-to-$\jpsi$ Transitions} 

\author{N.~E.~Adam}
\author{J.~P.~Alexander}
\author{K.~Berkelman}
\author{D.~G.~Cassel}
\author{V.~Crede}
\author{J.~E.~Duboscq}
\author{K.~M.~Ecklund}
\author{R.~Ehrlich}
\author{L.~Fields}
\author{R.~S.~Galik}
\author{L.~Gibbons}
\author{B.~Gittelman}
\author{R.~Gray}
\author{S.~W.~Gray}
\author{D.~L.~Hartill}
\author{B.~K.~Heltsley}
\author{D.~Hertz}
\author{L.~Hsu}
\author{C.~D.~Jones}
\author{J.~Kandaswamy}
\author{D.~L.~Kreinick}
\author{V.~E.~Kuznetsov}
\author{H.~Mahlke-Kr\"uger}
\author{T.~O.~Meyer}
\author{P.~U.~E.~Onyisi}
\author{J.~R.~Patterson}
\author{D.~Peterson}
\author{E.~A.~Phillips}
\author{J.~Pivarski}
\author{D.~Riley}
\author{A.~Ryd}
\author{A.~J.~Sadoff}
\author{H.~Schwarthoff}
\author{M.~R.~Shepherd}
\author{S.~Stroiney}
\author{W.~M.~Sun}
\author{D.~Urner}
\author{T.~Wilksen}
\author{M.~Weinberger}
\affiliation{Cornell University, Ithaca, New York 14853}
% ...
\collaboration{CLEO Collaboration} %FOR PRL,CLNS
\noaffiliation

\date{March 16, 2005}

\begin{abstract} 

We describe new measurements of the inclusive and exclusive
branching fractions for $\psi(2S)$
transitions to $J/\psi$ using $e^+e^-$~collision data collected
with the CLEO detector operating at CESR.
All branching fractions and ratios of branching fractions reported here
represent either the most precise measurements to date
or the first direct measurements.
Indirectly and in combination with other CLEO measurements,
we determine ${\cal B}(\chicJ\to\gamma J/\psi)$ and
${\cal B}[\psi(2S)\to {\rm light\ hadrons}]$. 

\end{abstract}

\maketitle

Heavy quarkonium states, non-relativistic bound 
$\ccbar$ or $b \bar b$ systems, 
offer a laboratory to study the strong interaction in the 
non-perturbative regime.
Charmonium in particular has served as a calibration tool
for the corresponding techniques and models~\cite{quarkoniumtheory}.  
The experimental situation for $\psi(2S)$~decays
has only begun to approach precisions at the percent level, 
with a global fit to the myriad of measurements from 
different experiments and eras 
revealing possible systematic inconsistencies~\cite{pdg2004}. 
Clarification of this picture is warranted.

This Letter presents branching fraction measurements 
of the four exclusive hadronic transitions
$\pp \to \jpsi + h$ ($h=\pi^+\pi^-,\ \pi^0\pi^0,\ 
\eta,\ \pi^0$), the exclusive channels 
$\pp \to \jpsi + \gamma\gamma$ through
$\pp \to \gamma \chicJ$, an inclusive measurement of
$\pp \to \jpsiplusany $, ratios between the above, 
and several derived quantities.
Multiple issues can be investigated with this data:
the observed discrepancy~\cite{besXJPsi} between 
${\cal B}(\pi^0\pi^0\jpsi) /{\cal B}(\pi^+\pi^-\jpsi)$
and the isospin-based expectation awaits corroboration;
$\pi^0\jpsi$ as an isospin-violating decay, when compared
with $\eta\jpsi$, helps constrain quark mass ratios~\cite{kuangetal};
the $\chicJ$ data offer access to the $\chicJ \to \gamma\jpsi$ rates 
in combination with the $\pp \to \gamma\chicJ$ branching
fractions~\cite{nPsiPrime}; 
confirmation of the transition $\psi(2S) \to \gamma \chi_{c0} 
\to \gamma\gamma\jpsi$~\cite{CBAL,bestwophoton}; 
and
the first direct constraint of ${\cal B}(\psi(2S) \to \mbox{light hadrons})$
using measurements from only one experiment. 

We use $e^+e^-$~collision data at and below the $\pp$~resonance,
$\sqrt s = 3.686\gev$ (${\cal L} = 5.86\invpb$) and
$\sqrt s = 3.670\gev$ (``continuum'' data, ${\cal L} = 20.46\invpb$),
collected with the CLEO detector~\cite{cleo} operating at the Cornell
Electron Storage Ring (CESR)~\cite{cesr}. 
The detector features a solid angle coverage of $93\%$ for
charged and neutral particles. 
The charged particle tracking system operates in a 1.0~T~magnetic field
along the beam axis and achieves a momentum resolution of
$\sim 0.6\%$ at momenta of $1\gev/c$. The CsI crystal
calorimeter attains
photon energy resolutions of $2.2\%$ for $E_\gamma = 1\gev$
and $5\%$ at $100\mev$.  

The $\jpsi$ is identified through its decay
to $\dimu$ or $\diel$, 
and we demand that
$m(\jpsi) \equiv m(\dilep) = 3.02 - 3.22\gev$. 
The ratios of calorimeter shower energy to track momentum, $E/p$, 
for the lepton candidates, taken to be the two 
tracks of highest momentum in the event, 
must be larger than $0.85$ for one electron
and above $0.5$ for the other, or smaller than $0.25$ and 
below $0.5$ for muon pairs. 
In order to salvage lepton pairs that
have radiated photons and would hence fail
the $m(\jpsi)$ cut, we add bremsstrahlung photon candidates
found within a cone of 100$\,$mrad to the 
three-vector of each lepton track at the interaction point (IP).
For $\psi(2S) \to \jpsiplusany $, cosmic ray background is rejected 
based on the distance of the track impact
parameters to the~IP ($<2$mm), and
on the $\jpsi$~momentum ($p_{\jpsi}>50\mev/c$). 
Radiative lepton pair production 
and radiative returns to the $\jpsi$ are suppressed for this mode
by demanding $|\cos\theta_{\jpsi}|< 0.98$.

For the exclusive final states, requirements on momentum
and energy conservation are imposed: 
$( E_{\jpsi} + E_X ) / \sqrt s = 0.95 - 1.05$, 
$ | |p_{\jpsi}| - | p_X | | / \sqrt s < 0.07$.
For $\eta$ and single-$\pi^0$ transitions,
in which the $\jpsi$~is monochromatic, $p(\jpsi)$ must 
lie within $500-570\mev/c$ ($\pi^0$) or $150-250\mev/c$ ($\eta$).
Charged dipion transition candidates must have two tracks
of opposite charge lower in momentum than the lepton pair. 
We identify neutral pions through their decay into two photons.
Photon candidates must not align with the projection of any
track into the calorimeter.
We require $m(\gamma\gamma) = 90-170\mev$ for 
$\pi^0$~mesons in
$\pi^0\pi^0\jpsi$ and $\eta \to \pi^+\pi^-\pi^0$; 
stricter conditions are imposed in $\pi^0\jpsi$ to suppress background
from $\pp \to \jpsi\gamma\gamma$ through $\chicJ$:
$m(\gamma\gamma) = 110-150\mev$, and in addition a constraint 
that the decay not be too asymmetric. 
We find $\eta$~meson candidates through $\eta\to\gamma\gamma$ 
or $\eta \to \pi^+\pi^-\pi^0$ with $m(\gamma\gamma)$ or
$m(\pi^+\pi^-\pi^0) = 500-580\mev$. 
The $\pi^+\pi^-e^+e^-$ final state must have 
$m(\pi^+\pi^-) > 350\mev$ to suppress background 
from radiative Bhabha events with 
subsequent $\gamma \to e^+e^-$~conversion.
The invariant mass of the system recoiling against the $\pi^+\pi^-$
or $\pi^0\pi^0$ must lie inside $3.05 - 3.15\gev$. 
To reduce background from radiative transitions to $\chi_{c1,2}$
into $\pi^0\jpsi$ and $\eta (\to \gamma\gamma)\jpsi$, 
the least energetic
photon in the $\eta$ or $\pi^0$~candidate has to fulfill
$E_\gamma > 200\mev$; $E_\gamma = 30-100\mev$ is
also allowed for the $\pi^0$ in $\pi^0\jpsi$.
In general, photons must have $|\cos \theta_\gamma|_{\max} < 0.93$;
for $\pi^0\jpsi$, we require $|\cos \theta_\gamma|_{\max} < 0.8$ 
to suppress radiative lepton pair background with a
fake $\pi^0$.
Candidates for $\gamma\chi_{cJ}\to\gamma\gamma\jpsi$
are accepted if $p_\jpsi = 250-500\mev$
(to suppress background from
$\pi^0,\eta\jpsi$), the recoil mass from the two photons is 
within $3.05-3.13\gev$,
and the energy of the second most energetic photon, $\egammalow$, 
is within $90-150$, $145-200$, and $230-290\mev$ ($J=2,1$, and $0$).

Table~\ref{tab:Everything} displays for each mode the raw event
counts obtained with this selection as well as the
efficiency~$\epsilon$, which 
is determined from Monte Carlo~(MC) simulation
using the {\tt EvtGen} generator~\cite{evtgen} and a 
GEANT-based~\cite{geant} detector simulation together
with corrections based on the data.
The dipion invariant mass distribution
as produced by {\tt EvtGen} is slightly suppressed at high
and low $m(\pi\pi)$ to better match the data,
altering the efficiencies by $<0.5\%$.
The $\chi_{cJ}$ MC~samples use intrinsic widths from Ref.~\cite{pdg2004},
and angular distributions have been generated
according to the prescription in Ref.~\cite{chicJMC}. 
The $\jpsiplusany$ data sample is modeled by the sum of all
exclusive MC channels, weighted by their measured branching
fractions. 
The trigger efficiencies for all modes are measured using
a prescaled subset of candidates in each channel
that fulfilled much looser requirements. 

\begin{figure}
\caption{
For inclusively selected dimuon (left) and dielectron (right)
events, the distributions of the dilepton mass in the $\pp$~data (solid
circles), after subtraction of the luminosity-scaled continuum, 
and in MC (solid line).
The two peaks above 3.2~GeV in
the $m(\dimu)$ distributions correspond to backgrounds from 
$\chi_{c0,2}\to K^+K^-$ and $\chi_{c0}\to \pi^+\pi^-$. 
}
\label{fig:prl_inc}
\includegraphics*[width=6.5in]{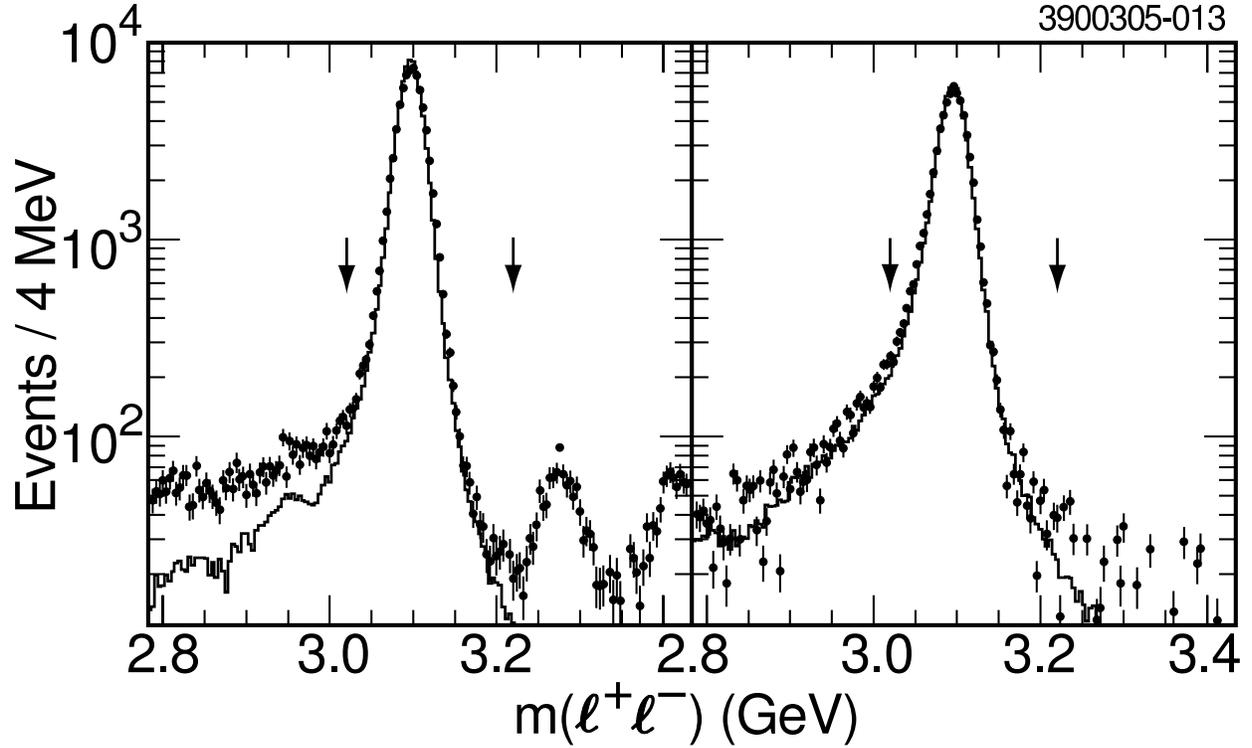}
\end{figure}

\begin{figure}
\caption{
For $\psi(2S)\to\pi^+\pi^-\ell^+\ell^-$ (left) and 
$\psi(2S)\to\pi^0\pi^0\ell^+\ell^-$ (right), $\diel$ and
$\dimu$ samples combined,
candidate events in the $\psi(2S)$
data (solid circles), MC simulation of signal (solid line),
and $\psi(2S)\to\eta J/\psi$ background (dashed histogram): 
distributions of the dilepton mass (top),
the mass recoiling against the dipion pair (middle),
and the invariant mass of the two pions.
}
\label{fig:prl_dipion}
\includegraphics*[width=6.5in]{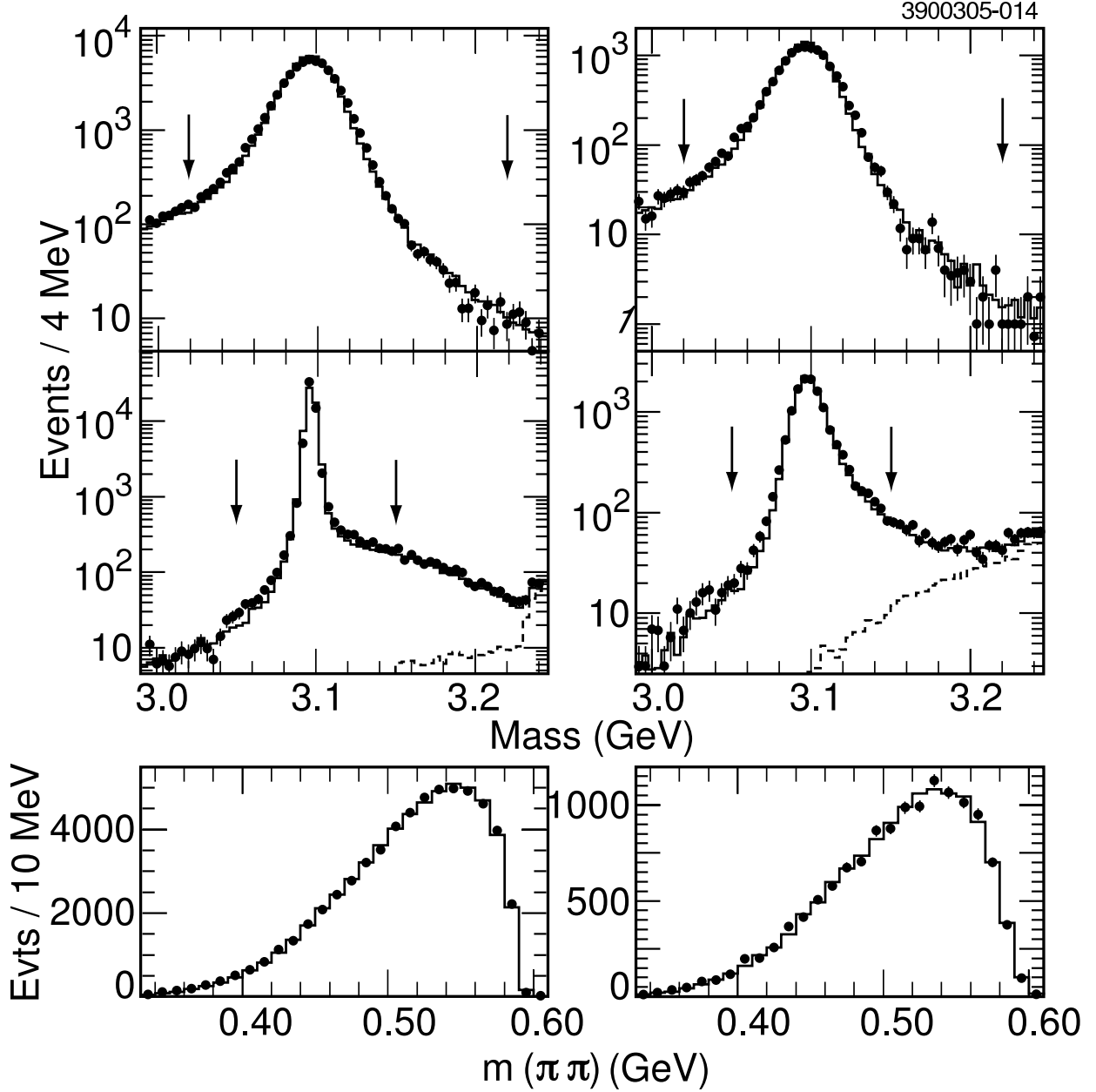}
\end{figure}

\begin{figure}
\caption{
For $\psi(2S)\to\eta(\to \gamma\gamma$, $\pi^+\pi^-\pi^0)\ell^+\ell^-$ 
(left) and
$\psi(2S)\to\pi^0\ell^+\ell^-$ (right), $\diel$ and
$\dimu$ samples combined,
candidate events in the $\psi(2S)$
data (solid circles) and MC simulation of signal (solid line):
distributions of the dilepton mass (top),
the $\jpsi$ momentum (middle),
and the invariant mass of the two photons.
In the lower left mass plot, the solid circles (data)
and solid line (MC) apply to $\eta \to\gamma\gamma$ decays, and
the open circles (data) and the dashed histogram (MC)
to $\eta \to \pi^+\pi^-\pi^0$.
}
\label{fig:prl_etapi0}
\includegraphics*[width=6.5in]{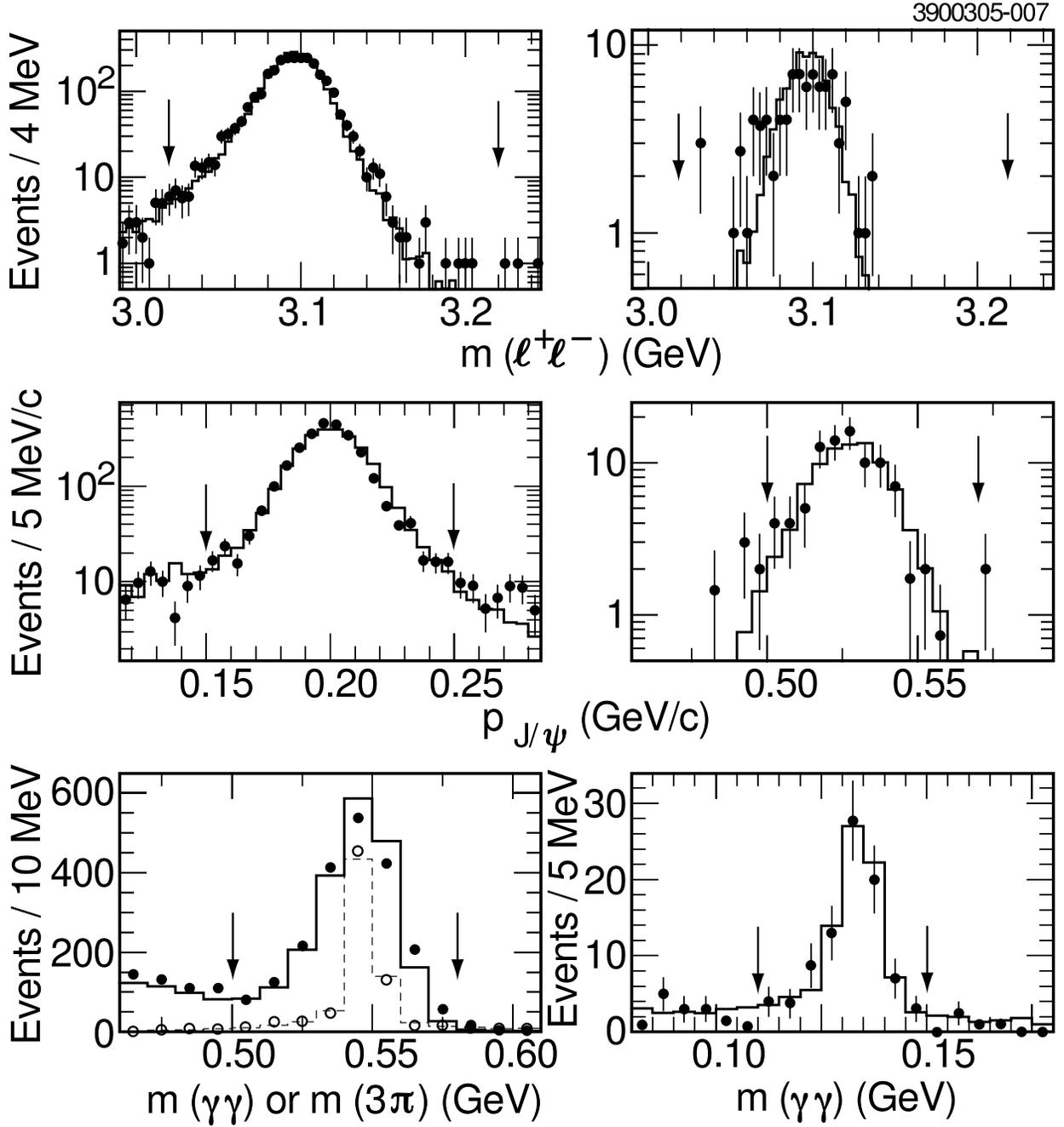}
\end{figure}

\begin{figure}
\caption{
For $\psi(2S)\to\gamma\chi_{cJ}$, $\chi_{cJ}\to\gamma J/\psi$,
$J/\psi\to\ell^+\ell^-$ candidate events in the $\psi(2S)$
data (solid circles) and MC simulation of signal (solid line), 
the distribution of the energy of the
second most energetic photon, $\egammalow$, (top), and the
two-photon recoil mass (bottom). The arrows
indicate nominal cut values. The inset offers a close-up of
the $\chi_{c0}$ region. The broken lines represent 
$\pi^0\pi^0\jpsi$~MC.
}
\label{fig:prl_chi}
\includegraphics*[width=6.5in]{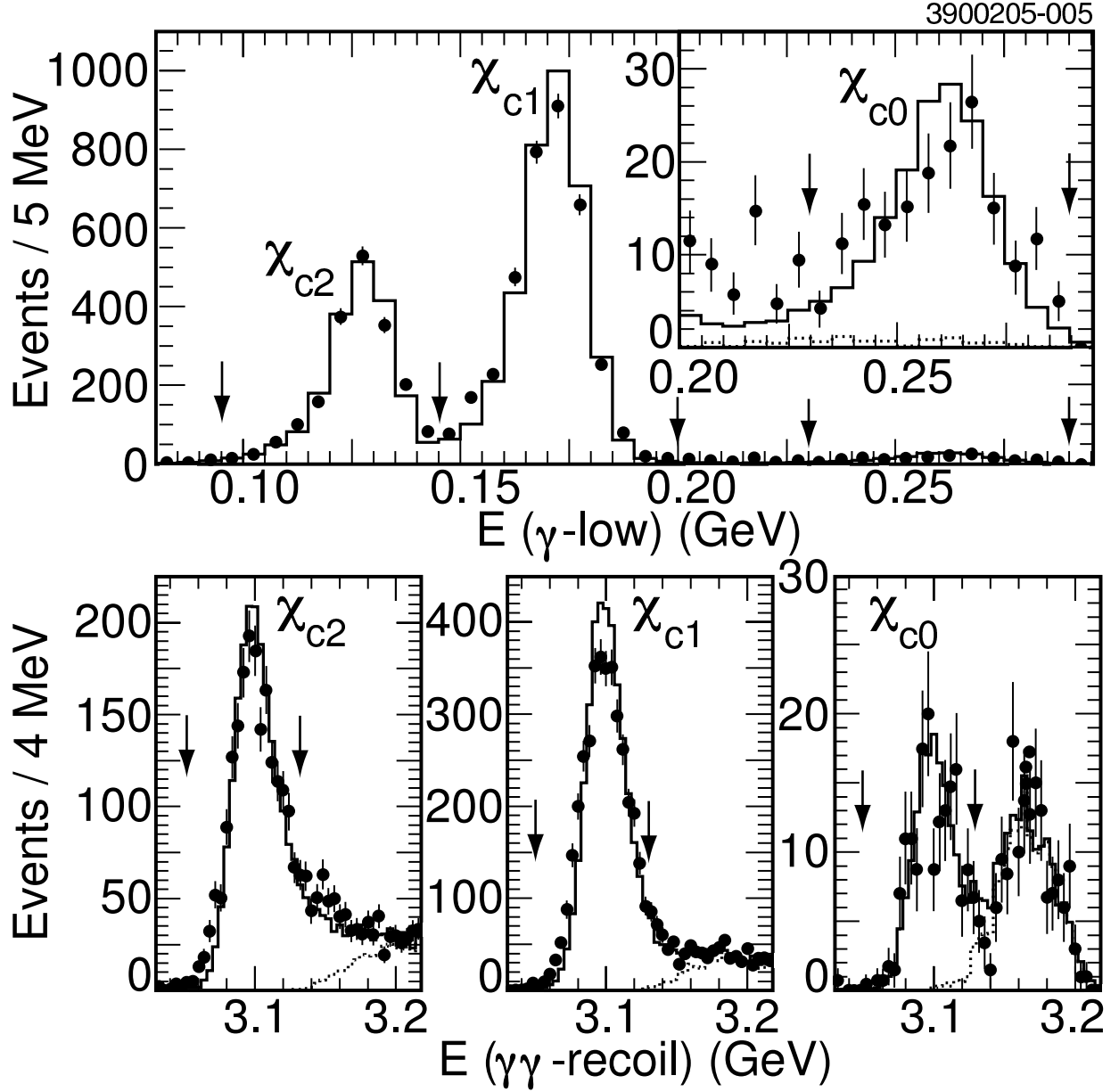}
\end{figure}

Data distributions of representative variables are shown in 
Figures~\ref{fig:prl_inc}-\ref{fig:prl_chi}
and compared to MC predictions.
All figures show distributions in which all selection criteria have
been applied to all variables {\sl except} for the one
shown. The MC~predictions in all figures depict the sum of all
exclusive channels; each source has been
normalized to our final branching fractions.
Distributions of invariant masses, angles, and momenta
show excellent agreement between MC and data for
all channels.

The observed event rates on the $\pp$ are corrected for 
contributions from continuum production and
$\pp$~cross-feed.
In all $\dimu$ and most $\diel$ modes, the observed
continuum yield is attributable to the Breit-Wigner tail of
the $\psi(2S)$.
The only significant $\pp$-induced backgrounds stem from 
cross-feed between the signal modes and from 
$\jpsi \to \pi^+\pi^-$ and $\rho\pi$. 
We estimate the sum of all contributions
to each channel from MC~simulation by determining for each
signal~MC what fraction passes the selection criteria 
of all other channels
relative to its own detection efficiency.
Cross-feed subtraction does not result in
a significant reduction of the event yield for most channels
(see Table~\ref{tab:Everything}). 
When analysis techniques similar to those in Refs.~\cite{CBAL,bestwophoton}
are applied to final states consisting of 
a~$\jpsi$ and two photons,
yields consistent with those presented here are obtained.

In order to measure the $\pi^\pm$, $\pi^0$, and lepton detection
efficiencies, we study $\psi(2S) \to \pi\pi\jpsi$, $\jpsi \to \dilep$ 
decays in which the selection of one pion (neutral or charged) or 
lepton is replaced by kinematic restrictions.
The samples thus obtained are very clean and give direct access to
the reconstruction efficiency of the not explicitly required, 
but usually
present, particle. We correct predicted MC efficiencies with
the observed, small MC-data discrepancies (all $\sim 1\%$ or less) found
in these studies and include them in the efficiencies in 
Table~\ref{tab:Everything}.
In the case of the dilepton selections, these corrections
absorb both any detector mismodeling and also that of
decay radiation~\cite{photos}. Relative systematic errors from these
studies are $0.75\%$ for each photon pair, $0.4\%$ per $\pi^\pm$,
$0.5\%$ per $\dimu$, and $0.2\%$ per $\diel$.
The uncertainty of lepton pair identification efficiency is $0.1\%$.

The systematic uncertainty stemming from 
cross-feed and background subtraction is a small
contribution to the total error, with the exception of 
$\pi^0\jpsi$ ($2.4\%$).
To account for potential mismodeling of the two-photon recoil
mass distribution, the $\gamma\gamma\jpsi$
channels are assigned an additional $2\%$ uncertainty. 
In the energy distribution of the second-most energetic
photon in $\gamma\chi_{cJ} \to \gamma \gamma\jpsi$ candidates, 
the data show an unexpected population 
in the region between the $\chi_{c1}$
and $\chi_{c0}$ (Figure~\ref{fig:prl_chi}). 
The events in $\egammalow=200-230\mev$
do not show any firm evidence for significant
contamination from continuum ($e^+e^-$~annihilation
not through a $\psi(2S)$), non-$\jpsi$ backgrounds,
anomalous levels of $\pi^0\jpsi$ or $\eta\jpsi$, or
an unmodeled, anomalously broad photon energy resolution,
although small fluctuations in all the sources mentioned are possible.
We cannot exclude that
these events originate at least partially from a high-side tail
of the $\gamma \chi_{c1}$ cascades not modeled by~MC, 
or as non-resonant $\gamma\gamma\jpsi$, or that
$\Gamma(\chi_{c0})=10.1\mev$~\cite{pdg2004} is an underestimate.
As no single source can be isolated and hence the
continuation of the background shape under the $\chi_{c0}$~peak
is unknown, we
apply an additional $10\%$ uncertainty for the $\chi_{c0}$~mode.
We add the above, 
the uncertainty on the $\jpsi\to\dilep$~branching fraction 
($1.2\%$~\cite{cleo_jpsidilep}), and $3\%$ as the estimate of
the precision of the number of $\pp$~decays~\cite{nPsiPrime},
all in quadrature. This last contribution dominates the systematic
error in the absolute branching fractions, 
with the exception of $\gamma\chi_{c0} \to \gamma\gamma\jpsi$.
Correlations between errors have been taken into account when
combining $\diel$ and $\dimu$ subsamples.
Many systematic uncertainties cancel in the ratios.

The branching fractions are readily obtained from the raw event yield
after background subtraction and correction for efficiency
by dividing by the number of
$\pp$~decays, $3.08 \times 10^6$, estimated by
the method described in Ref.~\cite{nPsiPrime}, and the 
$\jpsi \to \dilep$~branching fraction, 
$(5.953 \pm 0.056 \pm 0.042)\%$~\cite{cleo_jpsidilep}.
We also compute branching fraction ratios between
$\jpsiplusany$ and all exclusive modes and as well as
$\pi^+\pi^-\jpsi$ and all other exclusive modes. 
These results are included in Table~\ref{tab:Everything}.

Our investigation of the $\jpsi + h$ branching
fractions yields broad agreement with previous
results. 
The $\pi^0\pi^0\jpsi$ and $\eta (\to \pi^+\pi^-\pi^0)\jpsi$ 
measurements, along with many of the ratios of
branching fractions, are firsts of their kind.
The total errors 
match or improve upon current best
measurements~\cite{pdg2004,besXJPsi}. 
We observe the $\psi(2S) \to \gamma\chi_{cJ} \to \gamma\gamma\jpsi$ 
branching fractions to be 
slightly larger than BES~\cite{bestwophoton} and much larger
than CBAL~\cite{CBAL},
but the excellent agreement between our exclusive branching
fraction sum and the inclusive $\jpsi$ rate reinforces
the accuracy and internal consistency of this work.
We obtain $\Sigma {\cal B}[\pp \to \jpsi + h] 
+ \Sigma{\cal B}(\pp \to \gamma\chi_{cJ} \to \gamma\gamma\jpsi) = 
(58.9 \pm 0.2 \pm 2.0)\%$, 
consistent with ${\cal B} (\pp \to \jpsiplusany)$, thereby
not leaving much room, $(0.6 \pm 0.4)\%$, 
for other transitions to the $\jpsi$.
The branching fractions for transitions
through the $h_c$, 
$\pp \to \gamma\eta_c(2S) \to \gamma\gamma\jpsi$,
and $\pp \to \gamma\gamma\jpsi$ as a direct process
are not expected to exceed the observed difference.
 
These results enable us to calculate several 
derived quantities.
We measure the neutral and charged dipion branching
fraction to be consistent with the isospin-based expectation
of 1:2. 
The branching fraction for $\pp$ decaying to light hadrons,
computed as the difference between unity and
the branching fraction sum
of all exclusive direct transitions measured in this work
($\Sigma {\cal B}[\pp \to \jpsi + h ] = (53.4 \pm 0.2  \pm  1.7)\%$), 
the radiative decays~\cite{nPsiPrime} $\pp \to \chi_{cJ}\gamma$
and $\pp \to \gamma\eta_c$, and the dilepton branching 
fractions~\cite{pdg2004},
is found to be ${\cal B}[\pp \to \mbox{light hadrons}] =
(16.9 \pm 2.6)\%$. 
It can be compared with that of the $\jpsi$,
${\cal B}(\jpsi \to \mbox{light hadrons}) = 
(86.8 \pm 0.4)\%$~\cite{pdg2004,cleo_jpsidilep}, yielding a ratio of
$(19.4 \pm 3.0)\%$. 
Applying the ``12\%~rule''~\cite{exptwelvepercentrule} to inclusive
decays~\cite{guandli}, the ratio is $\sim 2.2\sigma$ above 
${\cal B}[\pp \to \dilep]/{\cal B}(\jpsi \to \dilep) = 
(12.6 \pm 0.7)\%$~\cite{pdg2004,cleo_jpsidilep}.
Combining the doubly-radiative branching fractions
analyzed in this study with those for
$\pp \to \gamma\chi_{cJ}$~\cite{nPsiPrime}, we arrive at
${\cal B}(\chi_{c0} \to \gamma\jpsi) = (2.0 \pm 0.2 \pm 0.2)\%$, 
${\cal B}(\chi_{c1} \to \gamma\jpsi) = (37.9 \pm 0.8 \pm 2.1)\%$, 
and
${\cal B}(\chi_{c2} \to \gamma\jpsi) = (19.9 \pm 0.5 \pm 1.2)\%$,
significantly higher than previous measurements for $J=0,1$. 
We measure the branching fraction ratio
${\cal B} [\pp \to \pi^0 \jpsi] / {\cal B} [\pp \to \eta \jpsi] 
= (4.1 \pm 0.4 \pm 0.1) \% $, to be compared with 
predictions ranging from $1.6\%$ (\cite{bestwophoton} 
based on Ref.~\cite{miller}) to 3.4\%~\cite{kuangetal}.

In summary, we have determined the branching fractions for
all exclusive $\pp \to \jpsi + h$ ($h = \pi^+\pi^-$, $\pi^0\pi^0$,
$\eta$, $\pi^0$) and $\pp \to \gamma\chi_{cJ}\to\gamma\gamma\jpsi$
transitions, with a similar strategy applied to all channels. 
We obtain results for ${\cal B}[\psi(2S) \to \jpsi + h]$ that are
consistent with but more precise than previous measurements, where available,
and ${\cal B}[\psi(2S) \to \gamma \chi_{cJ}\to\gamma\gamma\jpsi]$
values both larger and more precise than previous measurements.
The analysis is complemented by a study of the inclusive mode 
$\pp \to \jpsiplusany$, the production rate of
which is seen to be consistent
with that of the sum of all expected exclusive contributions.
Ratios between the branching fractions as well as
results on ${\cal B} (\chi_{cJ} \to \gamma\jpsi)$  
are also tendered.

We gratefully acknowledge the effort of the CESR staff 
in providing us with excellent luminosity and running conditions.
This work was supported by the National Science Foundation
and the U.S. Department of Energy.

\clearpage

\begin{table*}[htp]
\setlength{\tabcolsep}{0.4pc}
\catcode`?=\active \def?{\kern\digitwidth}
\caption{
For each mode: 
The detection efficiency, $\epsilon$, in percent; 
the numbers of events found in the $\psi(2S)$ and continuum samples, 
$N_{\psi(2S)}$ and $N_{\mathrm{cont}}$; 
the number of $\psi(2S)$ related background events, $N_{\mathrm{bgd}}$;
the branching fraction in percent 
and its ratio to ${\cal B}_{\jpsiplusany}$ and ${\cal B}_{\pi^+\pi^-\jpsi}$,
also in percent. 
}
\label{tab:Everything}
\footnotesize
\begin{center}
\hspace*{-7mm}
\begin{tabular}{l|cccc|c|c|c} 
 \hline 
 Channel & 
 $\epsilon$ & $N_{\pp}$ & $N_{\mathrm{cont}}$ & $N_{\mathrm{bgd}}$ & 
${\cal B}$ & ${\cal B}/{\cal B}_{X\jpsi}$ & ${\cal B}/{\cal B}_{\pi^+\pi^-\jpsi}$ \\  
 \hline 

          \PiPiJ
 & 49.3 
 &   60344
 &     221 
 &     113  
 & $ 33.54 \pm 0.14 \pm 1.10 $ 
 & $ 56.37 \pm 0.27 \pm 0.46 $
 & 
  \\ 
 \hline 

        \PizPizJ
 & 22.2
 &   13399
 &      67
 &     115
 & $ 16.52 \pm 0.14 \pm 0.58 $
 & $ 27.76 \pm 0.25 \pm 0.43 $ 
 & $ 49.24 \pm 0.47 \pm 0.86 $
  \\ 
 \hline 

     $\eta\jpsi$
 &  22.6 &     2793 & 17 & 116
 & $  3.25 \pm 0.06 \pm 0.11 $ 
 & $  5.46 \pm 0.10 \pm 0.07 $ 
 & $  9.68 \pm 0.19 \pm 0.13 $
  \\ 

 \quad 
         \EtaJGG
 & 16.9
 &    2065
 &      14
 &     103
 & $  3.21 \pm 0.07 \pm 0.11 $ 
 & $  5.39 \pm 0.12 \pm 0.06 $ 
 & $  9.56 \pm 0.21 \pm 0.14 $ 
  \\ 

 \quad 
    \EtaJThreePi
 &  5.8 
 &     728
 &       3
 &      13  
 & $  3.39 \pm 0.13 \pm 0.13 $
 & $  5.70 \pm 0.21 \pm 0.13 $
 & $ 10.10 \pm 0.38 \pm 0.22 $
  \\ 

 \hline 
           \PizJ
 & 13.9 
 &      88
 &       3
 &      20
 & $  0.13 \pm 0.01 \pm 0.01 $ 
 & $  0.22 \pm 0.02 \pm 0.01 $ 
 & $  0.39 \pm 0.04 \pm 0.01 $
  \\ 
 \hline 

       \ChicZero
 & 23.4
 &     172 
 &      20
 &      17
 & $  0.18 \pm 0.01 \pm 0.02 $ 
 & $  0.31 \pm 0.02 \pm 0.03 $
 & $  0.55 \pm 0.04 \pm 0.06 $
  \\ 
 \hline 

        \ChicOne
 & 30.6 
 &    3688 
 &      46
 &      21
 & $  3.44 \pm 0.06 \pm 0.13 $ 
 & $  5.77 \pm 0.10 \pm 0.12 $ 
 & $ 10.24 \pm 0.17 \pm 0.23 $
  \\ 
 \hline 

        \ChicTwo
 & 28.6
 &    1915 
 &      56
 &      62
 & $  1.85 \pm 0.04 \pm 0.07 $ 
 & $  3.11 \pm 0.07 \pm 0.07 $
 & $  5.52 \pm 0.13 \pm 0.13 $
  \\ 
 \hline 

             \XJ
 & 65.3
 &  151138 
 &   37916 
 &     123 
 & $ 59.50 \pm 0.15 \pm 1.90 $ 
 & 
 & 
  \\ 
 \hline 
 \hline 
\end{tabular} 
\end{center}
\end{table*}

\clearpage

\end{document}